# Updated Measurements of [O III] 88 μm, [C II] 158 μm, and Dust Continuum Emission from a $z = 7.2$ Galaxy

Yi W. Ren[1], Yoshinobu Fudamoto[2,3], Akio K. Inoue[1,2], Yuma Sugahara[2,3], Tsuyoshi Tokuoka[1], Yoichi Tamura[4], Hiroshi Matsuo[3], Kotaro Kohno[5,6], Hideki Umehata[4,7], Takuya Hashimoto[8], Rychard J. Bouwens[9], Renske Smit[10], Nobunari Kashikawa[6,11], Takashi Okamoto[12], Takatoshi Shibuya[13], and Ikkoh Shimizu[14]

[1] Department of Physics, Graduate School of Advanced Science and Engineering, Faculty of Science and Engineering, Waseda University, 3-4-1 Okubo, Shinjuku, Tokyo 169-8555, Japan; renyi@toki.waseda.jp
[2] Waseda Research Institute of Science and Engineering, Faculty of Science and Engineering, Waseda University, 3-4-1 Okubo, Shinjuku, Tokyo 169-8555, Japan
[3] National Astronomical Observatory of Japan, 2-21-1 Osawa, Mitaka, Tokyo 181-8588, Japan
[4] Department of Physics, Graduate School of Science, Nagoya University, Furocho, Chikusa, Nagoya 464-8602, Japan
[5] Institute of Astronomy, Graduate School of Science, The University of Tokyo, 2-21-1 Osawa, Mitaka, Tokyo 181-0015, Japan
[6] Research Center for the Early Universe, Graduate School of Science, The University of Tokyo, 7-3-1 Hongo, Bunkyo-ku, Tokyo 113-0033, Japan
[7] Institute for Advanced Research, Nagoya University, Furocho, Chikusa, Nagoya 464-8602, Japan
[8] Tomonaga Center for the History of the Universe (TCHoU), Faculty of Pure and Applied Sciences, University of Tsukuba, Tsukuba, Ibaraki 305-8571, Japan
[9] Leiden Observatory, Leiden University, NL-2300 RA Leiden, The Netherlands
[10] Astrophysics Research Institute, Liverpool John Moores University, 146 Brownlow Hill, Liverpool L3 5RF, UK
[11] Department of Astronomy, School of Science, The University of Tokyo, 7-3-1 Hongo, Bunkyo-ku, Tokyo, 113-0033, Japan
[12] Faculty of Science, Hokkaido University, N10 W8, Kitaku, Sapporo 060-0810 Japan
[13] Kitami Institute of Technology, 165 Koen-cho, Kitami, Hokkaido 090-8507, Japan
[14] Shikoku Gakuin University, 3-2-1 Bunkyocho, Zentsuji, Kagawa, 765-8505, Japan
Received 2022 May 11; revised 2023 January 20; accepted 2023 February 1; published 2023 March 8

## Abstract

We present updated measurements of the [O III] 88 μm, [C II] 158 μm, and dust continuum emission from a star-forming galaxy at $z = 7.212$, SXDF-NB1006-2, by utilizing Atacama Large Millimeter/submillimeter Array (ALMA) archival data sets analysed in previous studies and data sets that have not been analysed before. The follow-up ALMA observations with higher angular resolution and sensitivity reveal a clumpy structure of the [O III] emission on a scale of 0.32–0.85 kpc. We also combined all the ALMA [O III] ([C II]) data sets and updated the [O III] ([C II]) detection to $5.9\sigma$ ($3.6\sigma$–$4.5\sigma$). The non-detection of [C II] with data from the REBELS large program implies the incompleteness of spectral-scan surveys using [C II] to detect galaxies with high star formation rates (SFRs) but marginal [C II] emission at high-$z$. The dust continuum at 90 and 160 μm remains undetected, indicating little dust content of $< 3.9 \times 10^6 M_\odot$ ($3\sigma$), and we obtained a more stringent constraint on the total infrared luminosity. We updated the [O III]/[C II] luminosity ratios to $10.2 \pm 4.7$ ($6.1 \pm 3.5$) and $20 \pm 12$ ($9.6 \pm 6.1$) for the $4.5\sigma$ and $3.6\sigma$ [C II] detections, respectively, where the ratios in the parentheses are corrected for the surface brightness dimming effect on the extended [C II] emission. We also found a strong [C II] deficit (0.6–1.3 dex) between SXDF-NB1006-2 and the mean $L_{[\rm C\,II]}$–SFR relation of galaxies at $0 < z < 9$.

*Unified Astronomy Thesaurus concepts:* Galaxy formation (595); Galaxy evolution (594); Interstellar medium (847); High-redshift galaxies (734)

## 1. Introduction

During this decade, the Atacama Large Millimeter/submillimeter Array (ALMA) opens an atmospheric window for observing far-infrared (FIR) emission lines of oxygen, carbon, nitrogen, etc. arising from the interstellar medium (ISM) of distant galaxies existing in the Epoch of Reionization (EoR) and redshifted into the millimeter or submillimeter wavelength ranges when they reach the Earth. Emission lines emitted by ions or atoms in the ISM can be used to trace the star-forming processes and carry essential information about the properties of the ISM, like the gas density, metallicity, ionization parameter, etc. Dust, produced and destroyed by supernova explosions, is also relevant to star formation activities, and the continuum emission reemitted by dust after absorbing stellar radiation dominates the IR wavelength range. Therefore, both emission lines and dust continuum emission are important ingredients to understand the star formation of galaxies and galaxy formation and evolution (e.g., Hodge & da Cunha 2020).

Doubly ionized oxygen $\rm O^{2+}$, existing in H II regions, has well-known forbidden emission lines, among which optical lines like [O III] 4959/5007 were detected from galaxies up to $z > 9$ (e.g., Maiolino et al. 2008; Williams et al. 2022), while FIR fine-structure lines ([O III] 88 μm and [O III] 52 μm) were also detected from local galaxies with the Infrared Space Observatory (ISO), Herschel, and Stratospheric Observatory for Infrared Astronomy (SOFIA) (e.g., Malhotra et al. 2001; Negishi et al. 2001; Brauher et al. 2008; Madden et al. 2013; Spinoglio et al. 2022), and from galaxies at $z \sim 3$–4 using the Caltech Submillimeter Observatory (CSO) and Atacama Pathfinder Experiment (APEX) (e.g., Ferkinhoff et al. 2010; De Breuck et al. 2019). On the other hand, the optical lines are difficult to observe at higher redshift through ground-based telescopes due to atmospheric absorption, while FIR lines can be observed at $z > 6$ with ALMA (Inoue et al. 2014). Up to now, [O III] 88 μm emission has been detected in $>10$ $z \sim 6$–9 galaxies with ALMA (Inoue et al. 2016; Carniani et al. 2017; Laporte et al. 2017; Hashimoto et al. 2018, 2019; Tamura et al. 2019; Harikane et al. 2020; Witstok et al. 2022; Wong et al. 2022).

[C II] 158 $\mu$m emission, produced by singly ionized carbon $C^+$ and mainly arising from photodissociation regions (PDRs), H I clouds, etc., has also been observed at $z > 6$ galaxies. Even though this line was not detected from $z > 6$ Ly$\alpha$ emitters (LAEs) at the beginning of ALMA's operation (e.g., Ouchi et al. 2013; Ota et al. 2014), more ALMA observations successfully detected it from $z > 6$ galaxies after that (e.g., Maiolino et al. 2015; Knudsen et al. 2016; Pentericci et al. 2016; Bradac et al. 2017; Smit et al. 2018; Matthee et al. 2019; Bouwens et al. 2022; Schouws et al. 2022). Some researches reported non-detection of [C II] 158 $\mu$m emission in several [O III] emitters (Inoue et al. 2016; Hashimoto et al. 2018; Laporte et al. 2019), resulting in very high [O III]/[C II] luminosity ratios which are close to or exceed the upper bound of those of local metal-poor dwarf galaxies investigated by the Hershel Dwarf Galaxy Survey and regarded as local analogs of high-z galaxies due to their low metallicities and young stellar populations (Madden et al. 2013; De Looze et al. 2014; Cormier et al. 2015). Meanwhile, [C II] luminosity as a function of the star formation rate (SFR) ($L_{[C II]}$–SFR) of $z > 6$ galaxies shows a steeper slope compared with local galaxies (e.g., Ouchi et al. 2013; Ota et al. 2014; Knudsen et al. 2016; Bradac et al. 2017; Laporte et al. 2019; Harikane et al. 2020). Harikane et al. (2020) pointed out that the high [O III]/[C II] luminosity ratios and steeper $L_{[C II]}$–SFR relation of $z > 6$ galaxies may be caused by a high ionization parameter or low PDR covering fraction.

On the other hand, Schaerer et al. (2020) studied 118 normal star-forming galaxies at $z \sim 4$–6 from the ALPINE survey and concluded that the $L_{[C II]}$–SFR relation has no or little evolution from now to $z \sim 9$ when the SFRs are updated by themselves. Carniani et al. (2018) reanalysed $z \sim 5$–7 star-forming galaxies with ALMA archival data and also argued that the $L_{[C II]}$–SFR relation at early epochs is consistent with the local universe even with large dispersion, when associating the [C II] and UV emission properly and taking into account [C II] multi-clumps. More notably, Carniani et al. (2020) investigated the surface brightness dimming (SBD) effect on [C II] detected with ALMA by reanalysing ALMA archival data at $z \sim 6$–9 and performing simulations. They recovered [C II] detection from galaxies which had previously reported non-detections by combining observations with low angular resolution and applying a uv-taper when imaging. Their simulations also showed that the SBD effect would cause a flux loss of about 20%–40% even when the angular resolution is comparable with the source size. After modifying the [C II] detection and correcting for the SBD effect, Carniani et al. (2020) found that the [O III]/[C II] luminosity ratios of $z \sim 6$–9 galaxies are more consistent with local dwarf galaxies and the $L_{[C II]}$–SFR relation does not, or evolves little, across cosmic time.

Motivated by this controversial situation, we targeted an LAE, SXDF-NB1006-2, in this paper. This galaxy is the first galaxy from which [O III] 88 $\mu$m emission was detected with ALMA at $z > 6$, and its redshift was spectroscopically confirmed at $z = 7.212$ by the [O III] 88 $\mu$m emission (Inoue et al. 2016). [C II] 158 $\mu$m and dust continuum emission was not detected by Inoue et al. (2016), resulting in an extremely high [O III]/[C II] luminosity ratio of >12 (3$\sigma$). Later, Carniani et al. (2020) reported a 4.1$\sigma$ detection of [C II] 158 $\mu$m emission from SXDF-NB1006-2 by analysing additional [C II] data with lower angular resolution and applying a uv-taper when imaging. With SBD-corrected [C II] detection, they also reported that the [O III]/[C II] luminosity ratio and $L_{[C II]}$–SFR relation of SXDF-NB1006-2 are more in line with the local relation within the dispersion. In this work, we reanalysed the ALMA archival data used in previous works, and we also combined it with additional ALMA archival data that has never been analysed before to give updated measurements of the [O III] 88 $\mu$m, [C II] 158 $\mu$m, and dust continuum emission from SXDF-NB1006-2. We also updated the [O III]/[C II] luminosity ratios and the $L_{[C II]}$–SFR relation of SXDF-NB1006-2 with our updated measurements.

This paper is organized as follows. In Section 2, we describe our data and analysis of the [O III] 88 $\mu$m, [C II] 158 $\mu$m, and dust continuum emission obtained from multiple ALMA observations. We present the results in Section 3 and discussions in Section 4. The last section is devoted to the conclusions. In this paper, we assume $H_0 = 70$ km s$^{-1}$ Mpc$^{-1}$, $\Omega_M = 0.3$, $\Omega_\Lambda = 0.7$, $M_\odot = 2.0 \times 10^{30}$ kg, and $L_\odot = 3.83 \times 10^{26}$ W, where appliable. In this case, the angular size of 1″ is equivalent to 5.134 kpc at $z = 7.212$.

## 2. Data

### 2.1. Observations

We used two observations of [O III] 88 $\mu$m emission (Project IDs 2013.1.01010.S and 2015.A.00018.S) and three observations of [C II] 158 $\mu$m emission (Project IDs 2012.1.00374.S, 2013.A.00021.S, and 2019.1.01634.L). Detailed information about these five data sets is summarized in Table 1. The Cycle 7 [C II] 158 $\mu$m data set was taken from the REBELS large program (Bouwens et al. 2022).

### 2.2. Data Processing and Imaging

We obtained the calibrated measurement sets (MSs) from the East Asian ALMA Regional Center, who processed the raw data by running `scriptForPI.py` for all the observations. Then, we used the Common Astronomy Software Application (CASA) package (McMullin et al. 2007) to do further data processing and analysis.

We used the `split` task to pick out the fields containing the target galaxy and the `concat` task to combine different MSs. The Cycle 1 [C II] data set calibrated with CASA v.4.2.1 has visibility weights treated as per-spectral window (spw) values and initialized to unity, while the other [C II] data sets calibrated with CASA v.4.5.2 or a later version have visibility weights treated as per-channel values and proportional to $1/\sigma_{ij}^2$, where $\sigma_{ij}$ is the sensitivity of the specific visibility.[15] To combine the image data from different cycles accurately, one needs to ensure that the visibility weights of all data are proportional to $1/\sigma_{ij}^2$;[16] thus, we used the `statwt` task to adjust the visibility weights of the Cycle 1 [C II] data set before combining it with data from later cycles. The `statwt` task estimates the sensitivity per visibility and then corrects the weights. We input all the spws in the `fitspw` parameter while excluding channels containing the serendipitous CO ($J = 5$–4)

---
[15] https://casa.nrao.edu/Memos/CASA-data-weights.pdf
[16] https://casaguides.nrao.edu/index.php/DataWeightsAndCombination

**Table 1**
Summary of the Observations Used in This Work

|  | Cycle Number | Project ID | Angular Resolution[a] | $\sigma$[b] mJy beam$^{-1}$ |
|---|---|---|---|---|
| [O III] 88 $\mu$m | 2 | 2013.1.01010.S | $0\rlap.{''}45 \times 0\rlap.{''}38$ | 0.24 |
|  | 3 | 2015.A.00018.S | $0\rlap.{''}14 \times 0\rlap.{''}13$ | 0.14 |
| [C II] 158 $\mu$m | 1 | 2012.1.00374.S | $0\rlap.{''}8 \times 0\rlap.{''}6$ | 0.11 |
|  | 3 | 2013.A.00021.S | $1\rlap.{''}9 \times 1\rlap.{''}0$ | 0.15 |
|  | 7 | 2019.1.01634.L | $1\rlap.{''}5 \times 1\rlap.{''}2$ | 0.16 |

**Notes.**
[a] Beam sizes of the moment-0 maps with natural weighting.
[b] rms of cube images with a bin width of 100 km s$^{-1}$ and created by using spws covering line frequency only.

emission from a galaxy at $z_{\rm phot} = 1.48$ (Williams et al. 2009), SSTSL2 J021856.14-051951.4.

Then, we used the `tclean` task to create cubes and continuum images. To make image cubes, we determined the redshifted frequencies at 0 km s$^{-1}$ of [O III] 88 $\mu$m and [C II] 158 $\mu$m emission by considering a redshift of 7.212, which is determined by the [O III] 88 $\mu$m emission reported by Inoue et al. (2016), and their rest-frame frequencies, 3393.006 GHz and 1900.54 GHz, respectively. The bin width of the [O III] 88 $\mu$m emission is 20 km s$^{-1}$, which is the same as used by Inoue et al. (2016), and that of the [C II] 158 $\mu$m emission is 25 km s$^{-1}$. We used natural weighting for both cube and continuum images. Because our target is an extragalactic object, we defined the velocities in the barycentric frame by setting the outframe parameter to be "BARY," and the velocity type to be the optical velocity by setting the veltype parameter to be "optical." Regarding the MS of the Cycle 3 [C II] data set, we found that in the sideband covering the line frequency, spws covering the line frequency are set in frequency division mode (FDM) for spectral line observations, while the other spws which do not cover the line frequency are set in time division mode (TDM) for continuum observations. Therefore, when creating the data cubes of the Cycle 3 [C II] observation, we only used the FDM spws covering the line frequency. We also checked the point-spread functions (PSFs) of the Cycles 2 + 3 [O III] and Cycles 1 + 3 + 7 [C II] data cubes. Both of them are well-distributed 2D Gaussians with the same shape as the beams.

### 2.3. Emission Line Images

From the cube images, we made integrated intensity (moment-0) maps by using the `immoments` task.

In Inoue et al. (2016), the integrated velocity range for the [O III] 88 $\mu$m emission is from $-300$ to $+230$ km s$^{-1}$, which corresponds to the observed frequency range of 413.59–412.86 GHz. In order to cover this frequency range, we set the integrated velocity range for the Cycle 3 [O III] data to be $-310$ to $+250$ km s$^{-1}$. The integrated velocity range for the combined [O III] data of Cycle 2 and Cycle 3 is $-280$ to $+260$ km s$^{-1}$, which allowed us to obtain a detection with a higher signal-to-noise ratio (S/N).

For the [C II] 158 $\mu$m emission, the integrated velocity width for all the data sets is 225 km s$^{-1}$. In detail, we integrated from $-80$ to $+145$ km s$^{-1}$ for the Cycle 1, Cycles 1+3, Cycle 7, and Cycles 1 + 3 + 7 [C II] data sets, and from $-70$ to $+155$ km s$^{-1}$ for the Cycle 3 [C II] data set. When an atmospheric ozone line at $\sim$231.28 GHz was included in the integrated velocity range, we masked the corresponding channels. In addition, there is a suspicious spike lying on the channel bin from $+120$ to $+145$ km s$^{-1}$ in the spectrum of the Cycle 1 cube image, which causes a flux boost on the integral frequency edge on the spectrum of the Cycles 1 + 3 + 7 [C II] cube image. In this case, we also tried an integral of $-80$ to $+120$ km s$^{-1}$ for the Cycles 1 + 3 + 7 [C II] data without this bin. See Section 3.2.2 for our discussion.

### 2.4. Continuum Images

We made continuum images by using the Cycles 2 + 3 [O III] data and Cycles 1 + 3 + 7 [C II] data to measure the continuum emission detected by Band 8 and Band 6, respectively. To make continuum images, we excluded channels covering the integral frequency ranges of the Cycles 2 + 3 [O III] and Cycles 1 + 3 + 7 [C II] data to avoid influence from any line flux.

### 2.5. Flux Measurements

We used the `imfit` task to measure the integrated fluxes and beam deconvolved source sizes (FWHM) by fitting 2D Gaussian functions to sources in the images. We also used the `specfit` task to measure the FWHM velocities of the emission lines by fitting one or more 1D Gaussian functions to the spectrum extracted from a single pixel that has the largest brightness, i.e., representing a one-beam region.

In the case of non-detection, we provide 3$\sigma$ upper limits to the flux by measuring the background rms. To measure the rms of a data cube and moment-0 map, we used 3$\sigma$ clipping to measure the background noise better. To be specific, we measure the median and standard deviation of the image, then remove outliers out of the range of (median $\pm 3\sigma$). After that, we let this procedure iterate ten times to get a more accurate background noise level.

## 3. Results

### 3.1. [O III] 88 $\mu$m Emission

We combined observations with higher (Cycle 3) and lower (Cycle 2) angular resolutions and sensitivities to study the [O III] 88 $\mu$m emission from SXDF-NB1006-2. Figure 1 shows moment-0 maps of the [O III] 88 $\mu$m emission. The high-resolution Cycle 3 data reveal a clumpy structure of [O III] emission as shown in Panel (a). Panel (b) of Figure 1 shows a moment-0 map of the combined data without a uv-taper. Panel (c) shows the smooth 5.3$\sigma$ signal detected in the Cycle 2 [O III]

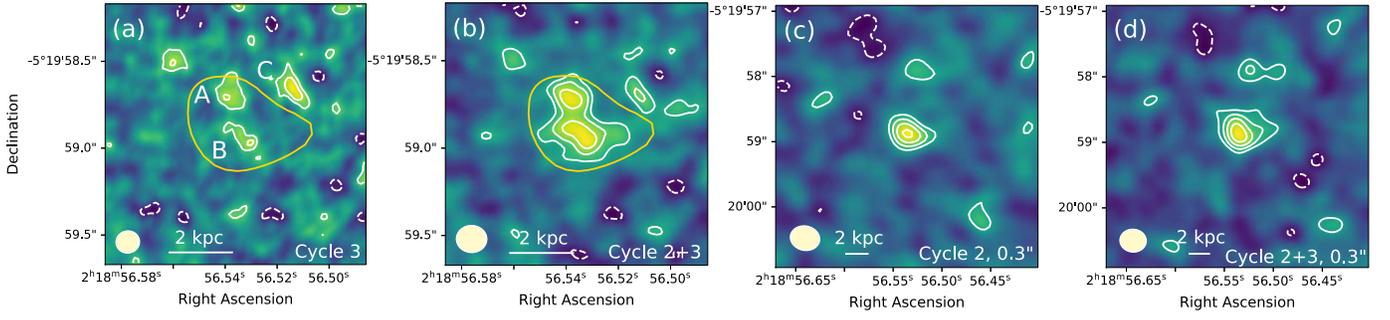

**Figure 1.** Integrated intensity maps of [O III] 88 μm emission. (a) Cycle 3 data without a uv-taper. (b) Cycle 2 and Cycle 3 combined data without a uv-taper. (c) Cycle 2 data with a 0″.3 uv-taper. (d) Cycle 2 and Cycle 3 combined data with a 0″.3 uv-taper. White solid lines indicate +2σ, +3σ, +4σ, and +5σ contours, and white dashed lines indicate −2σ contours. The yellow line in panels (a) and (b) is the 2σ contour of the Cycle 2 [O III] 88 μm detection (panel (c); see also Inoue et al. 2016). The image size of panels (a) and (b) is 1″.5 × 1″.5, and that of panels (c) and (d) is 4″ × 4″. The ellipses in the left bottom corners indicate the ALMA synthesized beam size.

**Table 2**
Summary of the S/Ns, Integrated Fluxes, and Beam Deconvolved Source Sizes Measured from the Cycle 2, Cycle 3 and Cycles 2 + 3 [O III] 88 μm Data

|  | S/N | Integrated Flux (Jy km s$^{-1}$) | Major/Minor Axis FWHMs (″) |
|---|---|---|---|
| Cycle 2 | 5.3/5.3[a] | 0.457 ± 0.140/0.45 ± 0.09[a] | 0.302 ± 0.195 × 0.158 ± 0.112 |
| Cycles 2 + 3 | 5.9 | 0.570 ± 0.151 | 0.523 ± 0.177 × 0.382 ± 0.183 |
| Cycle 3 A | 3.2 | 0.198 ± 0.090 | 0.173 ± 0.090 × 0.023 ± 0.087 |
| Cycle 3 B | 3.2 | 0.205 ± 0.104 | 0.247 ± 0.137 × 0.019 ± 0.092 |
| Cycle 3 C | 3.8 | 0.234 ± 0.096 | 0.330 × 0.083 |

**Note.**
[a] Values are from Inoue et al. (2016). The results of the Cycles 2 + 3 data are from an image with a 0″.3 uv-taper.

observation. Panel (d) shows a smoothed 5.9σ detection of the combined data after applying a uv-taper.

Table 2 summarizes the measurements of the signal in Cycle 2, three signals in Cycle 3, and the combined detection in Cycles 2 + 3 [O III] data, and Figure 2 shows single-beam spectra at the peak positions of them. The pink lines are best-fit Gaussian profiles and their velocity centers, intensity peaks, and FWHMs are summarized in Table 3.

*3.1.1. Line Detection and Total Flux*

To measure the total flux of the [O III] 88 μm emission line integrated over the whole galaxy, we created lower angular resolution images of the Cycles 2 + 3 [O III] data by applying a uv-taper. We applied a uv-taper using taper FWHMs from 0″.1 to 0″.5 with a step of 0″.1. We then searched for a taper FWHM that yielded the highest S/N of the [O III] line. We found the highest S/N of 5.9σ with a 0″.3 uv-taper (panel (d) of Figure 1). The resulting synthesized beam FWHM is 0″.40 × 0″.35, and the image rms is 0.0415 Jy beam$^{-1}$ km s$^{-1}$, showing an integrated flux of 0.570 ± 0.151 Jy km s$^{-1}$. The beam deconvolved source size is (0″.523 ± 0″.177) × (0″.382 ± 0″.183). The measurements are summarized in Table 2.

The spectrum of Cycles 2 + 3 [O III] data shows a broad and multicomponent line characteristic (Figure 2), and its properties are well consistent with those of the Cycle 2 [O III] data (Inoue et al. 2016).

*3.1.2. High-resolution [O III] 88 μm Image from the Cycle 3 Data*

To study the details of the [O III] 88 μm emission, we focused on the Cycle 3 data, which has higher angular resolution and sensitivity than that used by Inoue et al. (2016). In the moment-0 map shown in panel (a) of Figure 1, we found three marginal signals with S/Ns > 3σ near the center, denoted by A, B, and C. These multiple clumps show that the [O III] 88 μm emission has a clumpy structure. The yellow line is the 2σ contour of the Cycle 2 moment-0 map (panel (c) of Figure 1). Clumps A and B are enclosed within it, while C lies outside of it. This suggests that SXDF-NB1006-2 has, at least, two highly star-forming regions.

The S/Ns, integrated fluxes, and beam deconvolved source sizes of those three components are summarized in Table 2. From this table, the sum of the integrated fluxes of A and B is 0.403 ± 0.138 Jy km s$^{-1}$, and the sum of A, B, and C is 0.637 ± 0.168 Jy km s$^{-1}$. Both of them are consistent with the total integrated flux of the 5.9σ detection (Section 3.1.1) within the 1σ confidence level. Therefore, from the observed fluxes, we cannot conclude if signal C belongs to the total [O III] 88 μm emission. From Table 2, we also found that the size of the [O III] clumps is ∼0.32–0.85 kpc.

When we performed Gaussian fitting on signal A using the specfit task, we set the parameter ngauss to equal to 2, because of the spectral feature at the position of signal A. Then, we obtained two best-fit Gaussian singlets, implying that there may be two different velocity components existing at the region of signal A.

*3.1.3. Detailed Integration of the Cycle 3 Data*

None of the spectra of clumps A, B, or C can be fit by a single Gaussian profile of which the velocity center is right at 0 km s$^{-1}$. Therefore, we performed detailed integrals over velocity to get a better detection for each clump (see the detailed spectra in Figure 2). The integral ranges and measurements of all components are summarized in Table 4.

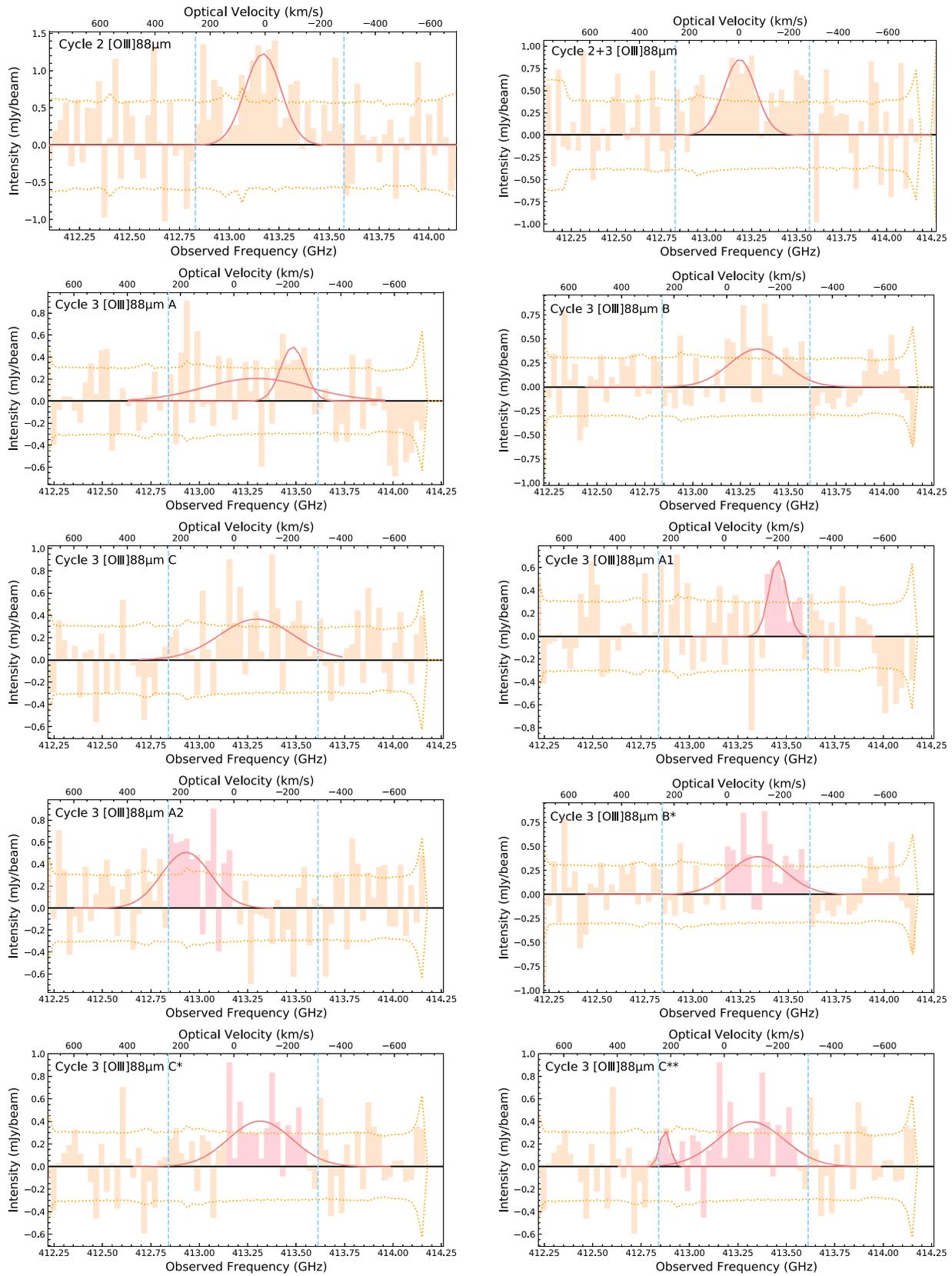

**Figure 2.** Spectra of [O III] 88 μm emission in the peak brightness pixel in the 0″.3 uv-taper images of Cycle 2 and Cycles 2 + 3 data (top row). The other panels show the spectra of the peak brightness pixels of the A, B, and C clumps (Section 3.1.2) and A1, A2, B⋆, C⋆, and C⋆⋆ clumps after detailed integrals of the Cycle 3 [O III] 88 μm data (Section 3.1.3). The blue dashed lines indicate the integral edges when we created the Cycle 2, Cycle 3, and Cycles 2 + 3 moment-0 maps. The pink-colored regions indicate the integral ranges of the detailed integrals. The pink solid lines are the best-fit Gaussian profiles of every clump. The orange dotted lines show the rms noise levels.

**Table 3**
Summary of the Properties of the Gaussian Fits

|  | Central Velocity[a] (km s$^{-1}$) | Peak Intensity (mJy beam$^{-1}$) | FWHM (km s$^{-1}$) |
| --- | --- | --- | --- |
| 0″.3 uv-taper |  |  |  |
| Cycle 2 | 7.7 ± 18.2 | 1.230 ± 0.230 | 156.6 ± 7.9 |
| Cycles 2 + 3 | −3.6 ± 18.8 | 0.850 ± 0.175 | 152.9 ± 9.0 |
| A | 98 ± 77 | 0.192 ± 0.116 | 268 ± 191 |
|  | −221 ± 18 | 0.490 ± 0.168 | 105 ± 43 |
| A1 | −201 ± 11 | 0.665 ± 0.200 | 77 ± 27 |
| A2 | 179 ± 30 | 0.507 ± 0.146 | 216 ± 70 |
| B | −116 ± 32 | 0.394 ± 0.102 | 250 ± 76 |
| B$^\star$ | −116 ± 32 | 0.394 ± 0.102 | 250 ± 76 |
| C | −85 ± 47 | 0.365 ± 0.104 | 329 ± 112 |
| C$^\star$ | −98 ± 38 | 0.403 ± 0.111 | 280 ± 92 |
| C$^{\star\star}$ | −100 ± 38 | 0.398 ± 0.106 | 290 ± 94 |
|  | 222 ± 18 | 0.328 ± 0.282 | 43 ± 44 |
| [C II] 4.5σ | 54 ± 38 | 0.252 ± 0.078 | 231 ± 87 |

**Note.**
[a] The zero velocity is set at $z = 7.212$ (Inoue et al. 2016).

From Figure 3 (a), clump A divides into two signals with different velocity components, denoted by A1 and A2. The signal at the same position with B but having a narrower velocity range is denoted by B$^\star$. In Figures 3(b) and (c), we tried two integral ranges for clump C and the resulting signals are denoted by C$^\star$ and C$^{\star\star}$.

From Table 4, the sum of the integrated fluxes of signals enclosed in the 2σ contour of the Cycle 2 [O III] detection (i.e., A1, A2, and B$^\star$) is 0.442 ± 0.104 Jy km s$^{-1}$, which is consistent with the measurement of the Cycle 2 [O III] detection within 1σ. In the case of considering signals outside the 2σ contour of the Cycle 2 detection, the sum of the integrated fluxes is 0.633 ± 0.124 Jy km s$^{-1}$ or 0.715 ± 0.146 Jy km s$^{-1}$, when adding C$^\star$ or C$^{\star\star}$, respectively. Both of them are consistent with the measurement of the Cycle 2 detection within 1 or 2σ. Again, we cannot conclude if all or only part of these marginal ∼4σ signals belong to the total [O III] 88 μm emission.

The single-beam spectra at the peak positions of A1, A2, B$^\star$, C$^\star$, and C$^{\star\star}$ can also be seen in Figure 2, and their velocity centers, peak intensities, and FWHMs are also summarized in Table 3. We set ngauss equal to 3 when performing the Gaussian fitting on signals C$^\star$ and C$^{\star\star}$ because of the multicomponent characteristics of their spectra, but we only obtained one Gaussian singlet for C$^\star$ and two for C$^{\star\star}$ in the overlapping integral velocity range (Figure 2 and Table 3).

In summary, even though the Cycle 2 [O III] data reveal a smooth one-component structure for the [O III] 88 μm emission in the image, the detailed structure of [O III] in SXDF-NB1006-2 is clumpy under the Cycle 3 ALMA observation with higher angular resolution and sensitivity. There is one clump, C (or C$^\star$, C$^{\star\star}$), lying outside the 2σ contour of the Cycle 2 [O III] detection, and the overall integrated flux increases by a factor of 1.2–1.6 compared with the Cycle 2 [O III] detection, but these measurements are consistent with each other at the 1 or 2σ confidence level.

**Table 4**
Summary of the Integral Ranges and Measurements of the Signals Detected After the Detailed Integrals

|  | Integral Range (km s$^{-1}$) | rms (Jy beam$^{-1}$ km s$^{-1}$) | S/N | Integrated Flux (Jy km s$^{-1}$) |
| --- | --- | --- | --- | --- |
| A1 | −290 ∼ −150 | 0.016 | 3.8 | 0.062 ± 0.029 |
| A2 | 10 ∼ +250 | 0.023 | 4.4 | 0.156 ± 0.066 |
| B$^\star$ | −310 ∼ +10 | 0.025 | 4.4 | 0.224 ± 0.075 |
| C$^\star$ | −270 ∼ +30 | 0.024 | 4.5 | 0.191 ± 0.068 |
| C$^{\star\star}$ | −270 ∼ +250 | 0.033 | 4.0 | 0.273 ± 0.102 |

We also plotted the Cycle 3 [O III] clumps on the Lyα emission captured by the NB1006 band of the Subaru telescope (Shibuya et al. 2012) (Figure 4). Clump C overlaps with the Lyα emission. However, the Lyα emission is usually extended to a larger scale than the ultraviolet (UV) continuum due to resonant scattering between Lyα photons and neutral hydrogen atoms (Dijkstra 2014). Therefore, even in this case, we still conclude that it is hard to determine if signal C (or C$^\star$, C$^{\star\star}$) is part of the [O III] 88 μm emission or not.

The James Webb Space Telescope (JWST) is going to observe optical [O III] emission coming from SXDF-NB1006-2 during the Cycle 1 observation. Thus, the clumpy structure of the [O III] 88 μm emission and the existence of clump C can be verified in the near future.

### 3.2. [C II] 158 μm Emission

#### 3.2.1. Line Detection

Moment-0 maps of the [C II] 158 μm emission are shown in Figures 5 and 6. Although the integrated frequency range is almost the same for each cycle data, the [C II] emission signal is very unstable as seen in Figure 5. We plotted the +2σ, +3σ, and +4σ contours of the Cycles 1 + 3 + 7 [C II] detection on the moment-0 maps of Cycle 1 and Cycle 7 [C II]. As Inoue et al. (2016) reported, there is no >3σ signal in the Cycle 1 [C II] moment-0 map (Figure 5(a)). However, the weak 2.8σ signal matches well with the contours of the Cycles 1 + 3 + 7 [C II] detection. It is likely that the extended [C II] emission is resolved out under the high angular resolution observation of Cycle 1 (0″.8 × 0″.6).

Furthermore, we can see a clear one-component emission only in the Cycle 3 data and the moment-0 map is very similar to that presented in Carniani et al. (2020). On the other hand, we cannot find any signal in the Cycle 7 data, although the angular resolution and sensitivity are very similar to those of Cycle 3. We will discuss this further in Section 4.1.

After we combined the Cycle 1 and Cycle 3 data, the S/N becomes 4.8 and we found two additional marginal signals denoted as SE and NW in panel (d) of Figure 5. The SE component can also be seen in the Cycle 1 [C II] moment-0 map (panel (a) of Figure 5) and an image made by the combination of all the data (panel (a) of Figure 6).

Finally, after we combined all the available data from Cycle 1, Cycle 3, and Cycle 7, we obtained a 4.5σ detection for the [C II] 158 μm emission. The final image also exhibits the clumpy structure of [C II] emission (panel (a) of Figure 6). Figure 7 shows single-beam spectra at the peak positions of the central signal and the tentative signals of SE and NW with optimized detections (see below) from the Cycles 1 + 3 + 7 combined data. Two bins overlaid by the hatched pattern are those affected by the atmospheric ozone line at 231.28 GHz and were excluded when we made the moment-0 maps.

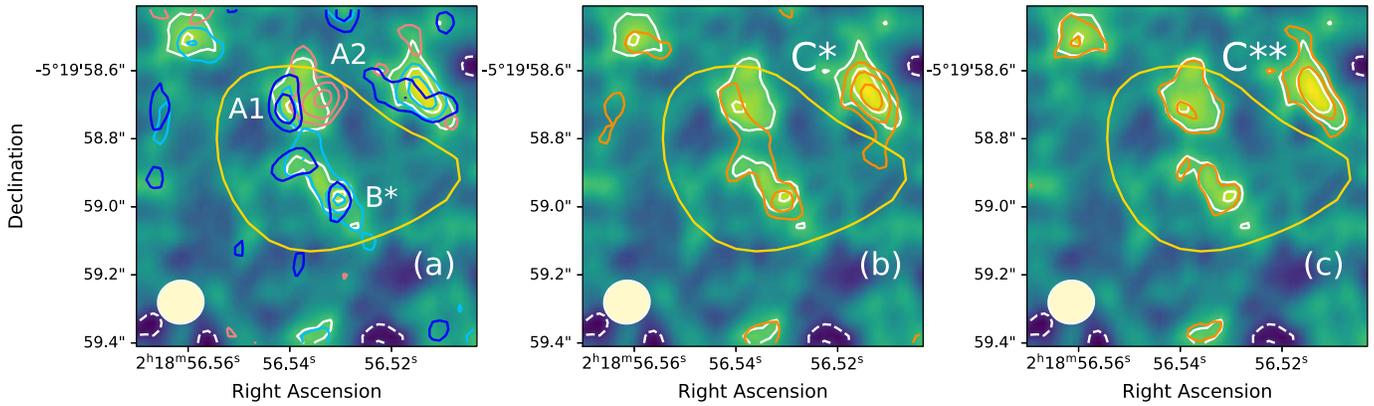

**Figure 3.** Integrated intensity maps of the Cycle 3 [O III] 88 μm emission with the results after detailed integrals overplotted. The solid lines indicate $+2\sigma$, $+3\sigma$, and $+4\sigma$ contours. The background images and white solid/dashed contours are the same as in panel (a) of Figure 1. Panel (a) shows signals A1, A2, and B* in deep blue, pink, and light blue colors, respectively. Panels (b) and (c) show signals C* and C** in orange color. The yellow line in every panel is the $2\sigma$ contour of the Cycle 2 [O III] 88 μm detection. The image size of every panel is $1'' \times 1''$. The ellipses in the left bottom corners indicate the size of ALMA's synthesized beams.

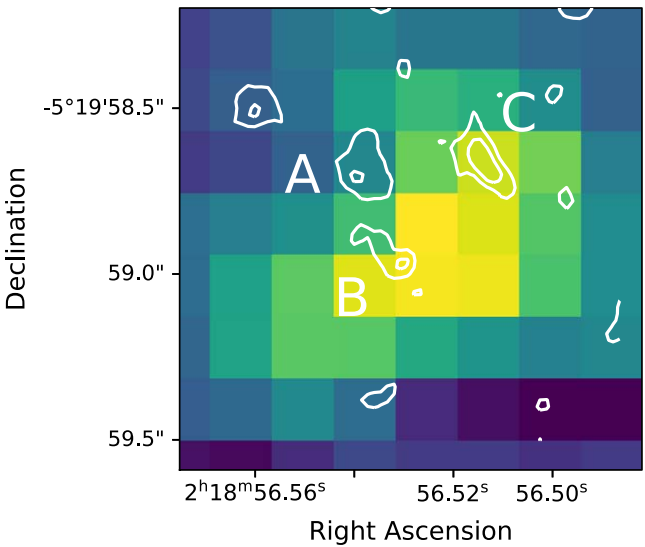

**Figure 4.** Image of the Lyα emission where the $2\sigma$ and $3\sigma$ contours of clumps A, B, and C of the Cycle 3 [O III] 88 μm data are overplotted. The size of the image is $1''\!.5 \times 1''\!.5$.

Measurements of the [C II] 158 μm emission are summarized in Table 5. For the Cycles $1+3+7$ [C II] detection, we found a beam deconvolved source size of $(1''\!.99 \pm 1''\!.04) \times (0''\!.09 \pm 0''\!.53)$ through 2D Gaussian fitting with the imfit task. The minor axis of the [C II] emission is unconstrained because the estimated minor axis is too small and its error is too large (Table 5). We also found a [C II]-based redshift of $z_{\rm [C\,II]} = 7.213 \pm 0.001$ from the velocity center given by the specfit task. The specfit result of the $4.5\sigma$ [C II] detection is listed in Table 3. The FWHM is $231 \pm 87 \, {\rm km \, s^{-1}}$, which is consistent with the former [C II] detection in Carniani et al. (2020).

By changing the integral velocity width, we optimized the detections of the SE and NW components (panels (b) and (c) of Figure 6). We first extracted single-beam spectra at the peak positions of the Cycles $1+3+7$ [C II] combined data for the SE and NW components (Figure 7). In the extracted spectra, we found that the positive signals come from different velocity ranges, so we made moment-0 maps for each component using slightly different channel ranges. We also took into account contamination from the atmospheric ozone line at $\sim 231.28$ GHz, so the integrated velocity ranges for SE and NW are $+70$ to $+170 \, {\rm km \, s^{-1}}$ and $-80$ to $+170 \, {\rm km \, s^{-1}}$, respectively. Then, we found the highest S/Ns of $3.2\sigma$ and $3.6\sigma$ for the SE and NW components, respectively.[17]

### 3.2.2. Possible Fluctuation of the [C II] Detection Signal

As we can see from the spectrum of the Cycles $1+3+7$ [C II] combined data (top-left panel of Figure 7), there is a channel bin that has a particularly high brightness ($>$2 times higher than the average of other bins) at the integral edge covering from $+120$ to $+145 \, {\rm km \, s^{-1}}$. In the bottom-right panel of Figure 7, we show the spectra of the Cycle 1, Cycle 3, Cycle 7, and Cycles $1+3+7$ [C II] data extracted from a region encompassing the $2\sigma$ contour of the Cycles $1+3+7$ [C II] 158 μm detection. From these spectra, we can see that the channel with a particularly high brightness in the Cycles $1+3+7$ [C II] spectrum is affected by a channel bin in the Cycle 1 spectrum with the same velocity range as the Cycles $1+3+7$ data.

To examine the robustness of our $4.5\sigma$ [C II] detection, we integrated the Cycle $1+3+7$ [C II] data from $-80 \, {\rm km \, s^{-1}}$ to $+120 \, {\rm km \, s^{-1}}$ (i.e., avoiding the exceptionally bright channel). The resulting S/N decreases to $3.6\sigma$ and the integrated flux decreases to $0.051 \pm 0.027 \, {\rm Jy \, km \, s^{-1}}$ (panel (d) of Figure 6), by a factor of $\sim 2$ smaller compared with the $4.5\sigma$ detection. In this case, the S/N of NW remains the same, while the significance of signal SE is less than $3\sigma$. Therefore, we conclude that the robustness of our $4.5\sigma$ [C II] detection is not strong, and we will use both the $4.5\sigma$ and $3.6\sigma$ detections to discuss the [O III]/[C II] luminosity ratio in Section 4.2, and the $L_{\rm [C\,II]}$–SFR relation in Section 4.3.

The imfit task cannot measure the beam deconvolved size of the $3.6\sigma$ [C II] well. Hence, we measured its size by measuring the major axis of the FWHM brightness contour on the image and then deconvolving the beam size from it. The resulting major axis of the beam deconvolved size is $1''\!.1$ and we will use this value for the discussion in Section 4.2 and Section 4.3. We did not measure the minor axis of the beam

---
[17] Our SE component also overlaps with the SE component offset from [O III] reported in Inoue et al. (2016) (see their Figure S4).

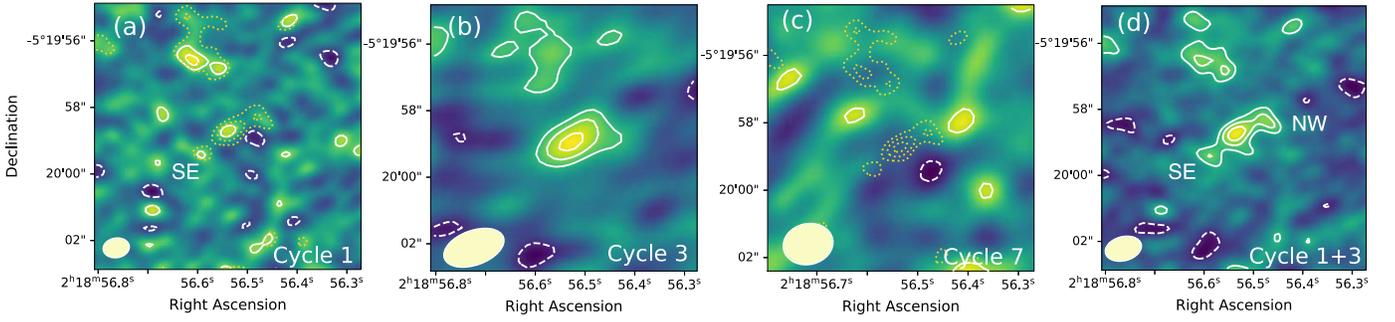

**Figure 5.** Integrated intensity maps of the [C II] 158 μm emission created by using (a) Cycle 1 data, (b) Cycle 3 data, (c) Cycle 7 data, and (d) Cycle 1 and Cycle 3 combined data. Apart from the signal located in the center, there may be another two components near the center, denoted by NW and SE. The white solid lines indicate +2σ, +3σ, and +4σ contours, and white dashed lines indicate −2σ contours. The yellow dotted lines in panels (a) and (c) indicate the +2σ, +3σ, +4σ contours of the Cycles 1 + 3 + 7 [C II] 158 μm detection. The image sizes are 8″ × 8″. The ellipses in the left bottom corners indicate the size of ALMA's synthesized beams.

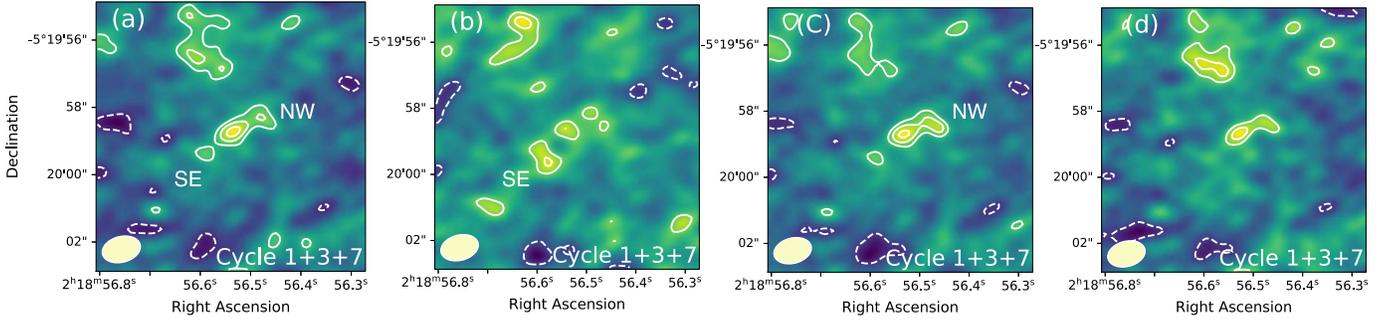

**Figure 6.** Same as Figure 5, but for the Cycles 1 + 3 + 7 combined data. Panels (b) and (c) illustrate SE and NW with their optimized S/Ns, respectively. Panel (d) shows the image obtained when integrating without one suspicious velocity bin.

**Table 5**
Summary of the Measurements of the Different Data Sets Targeting the [C II] 158 μm Emission

| | Beam Size | rms (Jy beam$^{-1}$ km s$^{-1}$) | S/N | Integrated Flux (Jy km s$^{-1}$) | Major/Minor Axis FWHM (″) |
|---|---|---|---|---|---|
| Cycle 1 | 0″8 × 0″6 | 0.017 | 2.8 | 0.022 ± 0.012 | ⋯ |
| Cycle 3 | 1″9 × 1″0 | 0.024 | 4.4 | 0.145 ± 0.053 | 0.89 ± 1.07 × 0.67 ± 0.31 |
| Cycle 7 | 1″5 × 1″2 | 0.028 | ⋯ | <0.088 (3σ) | ⋯ |
| Cycles 1+3 | 1″1 × 0″73 | 0.014 | 4.8 | 0.151 ± 0.051 | 2.10 ± 0.93 × 0.55 ± 0.44 |
| Cycles 1 + 3 + 7 | 1″2 × 0″77 | 0.013 (0.012)[a] | 4.5 (3.6)[a] | 0.100 ± 0.038 (0.051 ± 0.027)[a] | 1.99 ± 1.04 × 0.09 ± 0.53 |

**Notes.** "⋯" means non-detection.
[a] Results of integration without one peculiar bin at the integral edge. For the beam deconvolved size of the 3.6σ [C II] detection, please see Section 3.2.2.

deconvolved size because the minor axis in the image (panel (d) of Figure 6) seems smaller than the beam minor axis due to its low S/N.

### 3.3. Dust Continuum Emission

The dust continuum emission remains undetected, which is the same as the result of Inoue et al. (2016). The measurements of the dust continuum emission in Band 6 and Band 8 are summarized in Table 6. Compared with the previous work, we obtained better constraints on the flux density and total IR luminosity with the Band 6 observation. To calculate the total IR luminosity, we assumed the same dust temperature $T_{\rm dust}$ (40 K) and emissivity index $\beta$ (1.5) as Inoue et al. (2016), as well as using a modified blackbody function integrated over the 8–1000 μm wavelength range, without considering the effect of cosmic microwave background (CMB) temperature. In this case, we obtained a total IR luminosity of $L_{\rm TIR} < 4.3 \times 10^{10} L_\odot$ (3σ). To calculate the dust mass, we used Equation

**Table 6**
Measurements of the Dust Continuum Emission in Band 6 (1330 μm) and Band 8 (735 μm) and the Upper Limit on the Total IR Luminosity and Dust Mass

| $\lambda_{\rm obs}$ (μm) | rms (mJy beam$^{-1}$) | Flux Density (mJy) |
|---|---|---|
| 735 | 0.0235 | <0.071 (3σ) |
| 1330 | 0.0077 | <0.023 (3σ) |
| TIR luminosity ($L_\odot$) | <4.3 × 10$^{10}$ (3σ) | |
| Dust mass ($M_\odot$) | <3.9 × 10$^6$ (3σ) | |

(22) of Inoue et al. (2020) and the flux density in Band 6 with the same dust temperature and emissivity index as above. We also omitted the effect of the CMB temperature and assumed the mass absorption coefficient to be $\kappa_\nu = 30\,{\rm cm^2\,g^{-1}} \times (100\,\mu{\rm m}/\lambda)^\beta$ (Inoue et al. 2020).

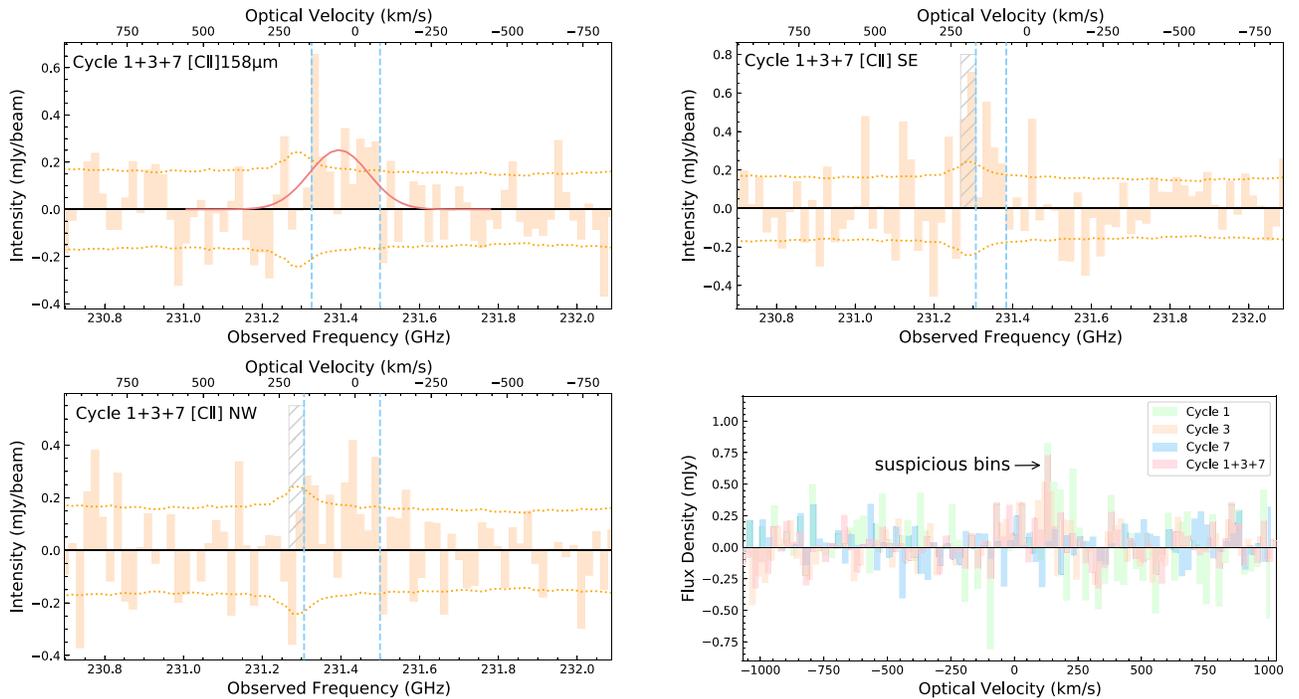

**Figure 7.** Spectra of [C II] 158 μm emission. The top-left, top-right, and bottom-left panels are single-beam spectra extracted from the peak position of the central signal and two possible components around it in the Cycles 1 + 3 + 7 [C II] moment-0 map. The two bins covered by the shaded patches are bins contaminated by the atmospheric ozone line at 231.28 GHz. The bottom-right panel shows spectra of the Cycle 1, Cycle 3, Cycle 7, and Cycles 1 + 3 + 7 [C II] data extracted from a region encompassing the 2σ contour of the Cycles 1 + 3 + 7 [C II] 158 μm detection.

## 4. Discussion

### 4.1. Non-detection of the Cycle 7 [C II] Observation

The Cycle 7 [C II] data set was obtained from the REBELS large program (2019.1.01634.L; PI: R. J. Bouwens), which aimed to detect the [C II] 158 μm or [O III] 88 μm emission lines from 40 $z > 6.5$ galaxies with the use of the spectral-scan strategy deployed on ALMA (Bouwens et al. 2022). The typical (requested) sensitivity limit for the [C II] luminosity in the REBELS large program is $2 \times 10^8 \, L_\odot$ ($3 \times 10^8 \, L_\odot$) for a 5σ detection (Bouwens et al. 2022). However, our final [C II] detection ranges from 3.6σ to 4.5σ with a luminosity range of $\sim(0.6$–$1.2) \times 10^8 \, L_\odot$. Thus, considering observations with sensitivity limits like those in the REBELS large program, our target is supposed to exhibit an ∼1.5σ–3σ (∼1σ–2σ) detection, which may be lower than the detection threshold (3σ) in this paper and may lead to non-detections.

Furthermore, the integration time of the Cycle 3 [C II] observation is 75.6 minutes. On the other hand, for the Cycle 7 data set analysed in this paper, there are three tunings scanning the frequency ranges covering the redshift likelihood distribution to detect [C II] emitter candidates in REBELS (REBELS-07; Bouwens et al. 2022), of which only one tuning with an integration time of 24 minutes covers the line frequency of SXDF-NB1006-2. Thus, the Cycle 7 observation has around only one-third of the integration time of the Cycle 3 observation.

However, in reality, the sensitivity of the Cycle 7 data is only slightly lower than that of the Cycle 3 data (Tables 1, and 5), because the sensitivity of an image also depends on other parameters apart from the integration time, such as the number of antennae, which in Cycle 7 (48) is larger than that of Cycle 3 (35) by a factor of 1.4. In conclusion, we cannot find out the reason why the [C II] emission was not detected with the Cycle 7 observation.

The REBELS large program found that more than ∼80% of their candidates with SFRs $>28 \, M_\odot \, \mathrm{yr}^{-1}$ can be detected at $\geqslant 7\sigma$ with the [C II] 158 μm emission in their project, showing that the spectral-scan strategy utilizing [C II] has very high efficiency in spectroscopically confirming galaxies at $z > 6.5$ (Bouwens et al. 2022). However, the [C II] emission of SXDF-NB1006-2 was missed by REBELS, implying the incompleteness of spectral-scan surveys like REBELS to detect galaxies at $z \sim 7$ with high SFRs but marginal [C II] emission. Instead, spectral-scan surveys using [C II] should consider the large dispersion of the $L_{[\mathrm{C\,II}]}$−SFR relation.

### 4.2. [O III]/[C II] Luminosity Ratio

The integrated fluxes and luminosities of the [O III] 88 μm and [C II] 158 μm emission are summarized in Table 7. For the [O III] 88 μm emission, we measured the integrated flux and luminosity by using the Cycles 2 + 3 combined data with a 0″.3 uv-taper, because it yielded the highest S/N of [O III] 88 μm emission (Section 3.1.1). For the [C II] 158 μm emission, we consider two cases for the integrated flux and luminosity: one is the 4.5σ detection, and the other one is the 3.6σ detection (Section 3.2.2). In the former case, we obtained an [O III] 88 μm/[C II] 158 μm luminosity ratio of $L_{[\mathrm{O\,III}]}/L_{[\mathrm{C\,II}]} = 10.2 \pm 4.7$, which is consistent with those of local dwarf galaxies within the uncertainty but close to the highest boundary (0.5–11; Madden et al. 2013; Cormier et al. 2015). While in the latter case, we obtained $L_{[\mathrm{O\,III}]}/L_{[\mathrm{C\,II}]} = 20 \pm 12$, which is also consistent with the local sample within the large uncertainty. These two ratios are consistent within the 1σ confidence level due to the large uncertainty of measurements for the 3.6σ detection. The open

Table 7
Measurements of the [O III] 88 μm and [C II] 158 μm Emission

|  | [O III] 88 μm[a] | [C II] 158 μm ($4.5\sigma$)[b] | [C II] 158 μm ($3.6\sigma$)[c] |
|---|---|---|---|
| Integrated Flux (Jy km s$^{-1}$) | $0.570 \pm 0.151$ | $0.100 \pm 0.038$ | $0.051 \pm 0.027$ |
| Flux (W m$^{-2}$) | $(7.9 \pm 2.1) \times 10^{-21}$ | $(7.7 \pm 2.9) \times 10^{-22}$ | $(3.9 \pm 2.1) \times 10^{-22}$ |
| Luminosity ($L_\odot$) | $(1.3 \pm 0.3) \times 10^9$ | $(1.2 \pm 0.5) \times 10^8$ | $(6.3 \pm 3.3) \times 10^7$ |

**Notes.**
[a] Values are from the Cycles 2 + 3 data with a 0″.3 uv-taper.
[b] Values are from the 4.5σ detection.
[c] Values are from the 3.3σ detection.

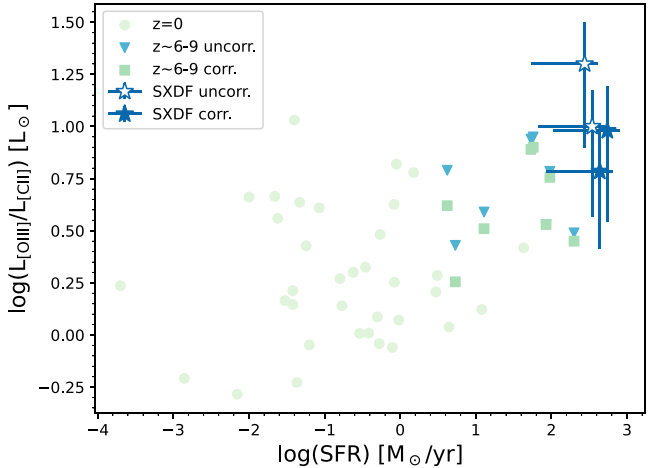

**Figure 8.** [O III]/[C II] luminosity ratios vs. SFR of this work (deep blue stars with errorbars), $z \sim 6$–9 galaxies uncorrected for the SBD effect (blue inverted triangles; Carniani et al. 2020; Harikane et al. 2020), $z \sim 6$–9 galaxies corrected for the SBD effect (green squares; Carniani et al. 2020; Harikane et al. 2020), and local dwarf galaxies from the Dwarf Galaxy Survey (light green circles; Madden et al. 2013; De Looze et al. 2014; Cormier et al. 2015). The three data points of this work were moved to the right by 0.1 dex (SBD-corrected 4.5σ–[C II]), 0.2 dex (SBD-corrected 3.6σ–[C II]), and to the left by 0.1 dex (SBD-uncorrected 4.5σ–[C II]), respectively, for display purposes.

deep blue star symbols in Figure 8 shows the observed [O III]/[C II] ratios compared with other samples.

Furthermore, we also measured $L_{[O III]}/L_{[C II]}$ after correcting for the SBD effect (Carniani et al. 2020). The major axis of the beam deconvolved size of the [C II] emission is (1″.99 ± 1″.04) and 1″.1 for the 4.5σ and 3.6σ detections, respectively. The source size $D_{source}$ in the simulations of Carniani et al. (2020) is the major axis of the FWHM of the 2D Gaussian profile, and the axial ratio and position angle of every simulated galaxy are randomly assigned by them. Therefore, in our calculation, we set the $D_{source}$ to be (1″.99 ± 1″.04) and 1″.1 for the 4.5σ and 3.6σ detections, respectively. The angular resolution of the Cycles 1 + 3 + 7 combined data is 1″.2 × 0″.77, so we set $\theta_{beam}$ to be $\sqrt{1″.2 \times 0″.77}$ in our calculation. Then, the angular resolution normalized by the intrinsic source size ($\theta_{beam}/D_{source}$) is $0.48 \pm 0.25$ and 0.87 for the 4.5σ and 3.6σ detections, respectively. In this case, we estimated the SBD correction factor to be $0.60 \pm 0.20$ and $0.48 \pm 0.12$ for the 4.5σ and 3.6σ detections from Figure 6 of Carniani et al. (2020), respectively. We have also considered the large uncertainties shown in Figure 6 of Carniani et al. (2020). Finally, we obtained [O III]/[C II] luminosity ratios after correcting for the SBD effect of $6.1 \pm 3.5$ and $9.6 \pm 6.1$ for the 4.5σ and 3.6σ

Table 8
A Summary of the [O III] 88 μm/[C II] 158 μm Luminosity Ratio

|  | [C II] 158 μm | |
|---|---|---|
|  | $4.5\sigma$ | $3.6\sigma$ |
| Observed [O III]/[C II] ratio | $10.2 \pm 4.7$ | $20 \pm 12$ |
| Beam size / [C II] size | $0.48 \pm 0.25$ | 0.87 |
| SBD correction factor[a] | $0.60 \pm 0.20$ | $0.48 \pm 0.12$ |
| Corrected [O III]/[C II] ratio | $6.1 \pm 3.5$ | $9.6 \pm 6.1$ |

**Note.**
[a] Estimated from the Figure 6 of Carniani et al. (2020).

detections, respectively. These numbers are also summarized in Table 8. The [O III]/[C II] luminosity ratios after correcting for the SBD effect are more consistent with those of local dwarf galaxies (Figure 8).

There are multiple papers discussing the reasons for the high [O III]/[C II] luminosity ratios of high-$z$ galaxies (e.g., Arata et al. 2020; Harikane et al. 2020), of which Katz et al. (2022) give a comprehensive discussion about the contributors to high [O III]/[C II] in high-$z$ galaxies through their cosmological simulations and comparisons with previous works. They consider that lower C/O abundance ratios, lower PDR covering fractions, and higher ionization parameters play important roles in contributing to high [O III]/[C II], while CMB attenuation, extended [C II] emission, inclination effects, and observational biases of high-$z$ galaxies are possible but less important. Simulations from Katz et al. (2022) also show that lower ISM densities contribute to high [O III]/[C II], but they also pointed out that the densities of regions emitting [O III] and regions emitting [C II] may not be the same, which is indeed confirmed by their simulations. Furthermore, even though their simulations do not show a correlation between metallicity and [O III]/[C II], they find a trend that decreasing metallicity is correlated with increasing ionization parameter and increasing [O III]/[C II] for galaxies with high SFRs. This suggests that lower metallicity for galaxies with high SFRs may also lead to high [O III]/[C II] luminosity ratios.

### 4.3. $L_{[C II]}$–SFR relation

Figure 9 shows a comparison between the results of SXDF-NB1006-2 and the $L_{[C II]}$–SFR relations from local H II/starburst galaxies ($\log(SFR/M_\odot\ yr^{-1}) = (1.00 \pm 0.04)\ \log(L_{[C II]}/L_\odot) - (7.06 \pm 0.33)$; De Looze et al. 2014), the ALPINE survey ($\log(L_{[C II]}/L_\odot) = (0.84 \pm 0.13)\ \log(SFR/M_\odot\ yr^{-1}) + (7.09 \pm 0.21)$; Schaerer et al. 2020), and $z > 6$ galaxies with modified [C II] detection and corrected for the SBD effect ($\log(L_{[\text{C II}]}/L_\odot) = (1.1 \pm 0.2) \log(\text{SFR}/M_\odot \text{ yr}^{-1}) + (6.8 \pm 0.2)$; Carniani et al. 2020). As for the $L_{[\text{C II}]}$−SFR relation of the ALPINE survey, we adopted the relation using $4 < z < 6$ star-forming galaxies, with SFRs derived from SED fitting, and $3\sigma$ upper limits for [C II] non-detections (Schaerer et al. 2020).

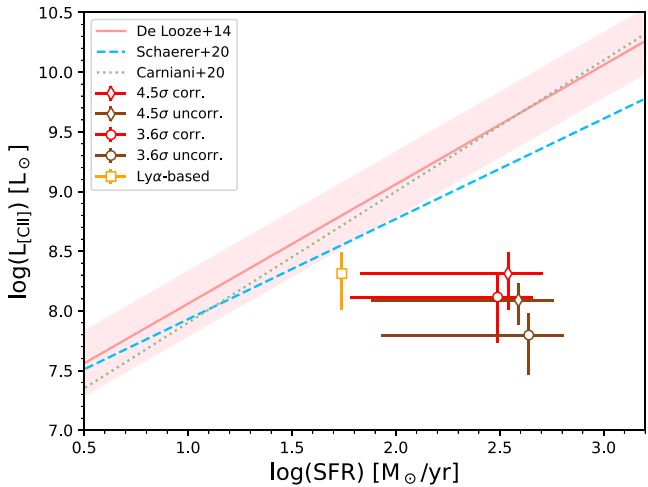

**Figure 9.** Our $L_{[\text{C II}]}$−SFR results for the $4.5\sigma$ (diamonds) and $3.6\sigma$ (circles) detections compared with the $L_{[\text{C II}]}$−SFR relations from local H II/starburst galaxies (pink line; De Looze et al. 2014), $4 < z < 6$ star-forming galaxies from the ALPINE survey (blue dashed line; Schaerer et al. 2020), and $z > 6$ galaxies with modified [C II] detections and corrected for the SBD effect (green dotted line; Carniani et al. 2020). The pink shaded region indicates a $1\sigma$ dispersion of 0.27 dex from De Looze et al. (2014). The red data points are the results after correcting for the SBD effect, while the brown points are uncorrected. All the red and brown data points have the same value ranges of logSFR, but we have moved the data points of $4.5\sigma$ SBD-uncorrected to the right by 0.05 dex, $3.6\sigma$ SBD-corrected to the left by 0.05 dex, and $3.6\sigma$ SBD-uncorrected to the right by 0.1 dex, for display purposes. The orange square is the Ly$\alpha$-based SFR corrected for the Ly$\alpha$ escape fraction. The value of the vertical axis for the orange square is from the $4.5\sigma$-[C II] corrected for the SBD effect. We omit the large uncertainty of the correction from Ly$\alpha$ escape fraction in the plot.

We plot SXDF-NB1006-2 using the SFR derived from SED fitting ($\log(\text{SFR}/M_\odot \text{ yr}^{-1}) = 2.54^{+0.17}_{-0.71}$, Table 1 of Inoue et al. 2016), and the [C II] luminosity measured from our $4.5\sigma$ and $3.6\sigma$ detections (red and brown data points, respectively). We also show the results before and after correcting for the SBD effect. We found that all the data points based on the SED-derived SFR are lying below the above three $L_{[\text{C II}]}$−SFR relations, showing a strong [C II] deficit. The [C II] luminosity of our results deviates from the $L_{[\text{C II}]}$−SFR relations by 0.6–1.3 dex.

There are two possibile reasons for the [C II] deficit in SXDF-NB1006-2. One possibile reason is the galaxy's intrinsic properties. From the SED measurements of Inoue et al. (2016), SXDF-NB1006-2 is inferred to have undergone drastic star formation over the most recent 1 Myr, hence the neutral gas budget where the [C II] emission mainly comes from is being consumed dramatically, leading to the [C II] deficit in the $L_{[\text{C II}]}$−SFR relation. Again, through the SED fitting results of Inoue et al. (2016), the escape fraction of ionizing photon of SXDF-NB1006-2 is estimated to be ∼50%. Meanwhile, the simulations from Katz et al. (2023) show that LyC leakers with an escape fraction of $f_{\text{esc}} > 20\%$ and having recent or past starburst activity at $z = 4.64$ and $z = 6$ will exhibit a [C II] deficit in the $L_{[\text{C II}]}$−SFR relation, where the SFR is averaged over the past 10 or 100 Myr, which gives us a clue for the reason for the [C II] deficit in SXDF-NB1006-2.

The other possible reason is the large uncertainty of measurements on the SFR (see also Section 4.4). We also plotted our result with the Ly$\alpha$-based SFR corrected for the Ly$\alpha$ escape fraction, which shows a weaker [C II] deficit than what we found when using the SED-derived SFR. Nevertheless, the uncertainty of the Ly$\alpha$ escape fraction is also large (Sobral & Matthee 2019). Future observations by JWST may solve this problem. JWST will observe the stellar components of SXDF-NB1006-2 during Cycle 1 with its excellent high angular resolution and sensitivity, thus being able to enhance the measurement of the SFR through SED fitting.

### 4.4. The Nature of SXDF-NB1006-2

Finally, we discuss the nature of SXDF-NB1006-2. According to the near-infrared (NIR) photometry of Inoue et al. (2016), this galaxy is detected in $J$ band with a magnitude of $25.46 \pm 0.18$ AB. But it is not detected in the $H$ and $K$ bands, resulting in a UV continuum slope of $\beta < -2.6$, where the flux density is $f_\lambda \propto \lambda^\beta$. This is extremely blue compared with galaxies at similar redshifts (Bouwens et al. 2014), so the SED fitting suggests a very young age of ∼1 Myr and zero dust attenuation (Inoue et al. 2016). Such a short duration of star formation predicts a faint UV continuum due to the small amount of stars. Nevertheless, this galaxy is detected in $J$ band, suggesting active star formation.

The Ly$\alpha$ luminosity of this galaxy was reported to be $(1.2^{+1.5}_{-0.6}) \times 10^{43}$ erg s$^{-1}$ (Shibuya et al. 2012). By using the $J$-band luminosity as a proxy for the Ly$\alpha$ continuum for simplicity, the Ly$\alpha$ equivalent width (EW) in the rest frame is estimated to be $47^{+59}_{-23}$ Å. If we calculate the Ly$\alpha$ continuum from the $J$-band continuum by assuming $\beta = -2.6$, the EW becomes $29^{+38}_{-14}$ Å. Thus, SXDF-NB1006-2 is securely classified as an LAE, which is defined to have an Ly$\alpha$ EW of >30 Å. The Ly$\alpha$ EW can be used as an indicator of the Ly$\alpha$ escape fraction (Sobral & Matthee 2019), yielding about 20%. Since the SFR from the observed Ly$\alpha$ luminosity is estimated to be 11 $M_\odot$ yr$^{-1}$ (Shibuya et al. 2012), the corrected SFR can be about 50 $M_\odot$ yr$^{-1}$. A large Ly$\alpha$ EW also indicates the efficient escape of Lyman continuum photons (i.e., ionizing photons) (e.g., Gazagnes et al. 2020). Indeed, Inoue et al. (2016) estimated ∼50% of the Lyman continuum escapes from SXDF-NB1006-2. Taking this into account, the true SFR could be SFR(Ly$\alpha$)/$f_{\text{esc,Ly}\alpha}/(1 - f_{\text{esc,LyC}}) \sim 100$ $M_\odot$ yr$^{-1}$, roughly consistent with the value estimated from the SED fitting.

Although there are some discussions that high-resolution observations made with ALMA can make a smooth disk look like clumpy (Gullberg et al. 2018), our ALMA observation of [O III] reveals a clumpy structure of the ionized gas in SXDF-NB1006-2. Multiple clumps with a diameter of 0.32–0.85 kpc are confined to a scale of ∼2.3 kpc. The [C II] emission also shows a clumpy structure. Figure 10 shows the spacial distribution of the Ly$\alpha$, [O III], and [C II] emission. From this figure, we can see that the [O III], [C II] and Ly$\alpha$ emission overlap with each other. The Ly$\alpha$ emission is spatially extended and envelops the [O III] clumps tracing the star-forming regions (see also Figure 4). The [C II] emission, which predominantly arises from H I gas, is further extended on a scale of ∼10 kpc. The phenomenon that the [C II] emission has a larger size than the [O III] emission can also be seen in local dwarf galaxies (Cormier et al. 2015). It is worth noting that from the top-right panel of Figure 10, the central component of [C II] encloses the [O III] clumps completely. Apart from the fact that the [C II] emission mainly arises from H I regions, which are larger than the ionized star-forming regions where the [O III] emission originates, it is also possible that SXDF-NB1006-2 is undergoing a galaxy merger event. The spatial extent of the Lyα emission may be explained by scattering with extended H I gas. However, if H I gas covers the ionized regions completely, the Lyman continuum escape becomes inefficient. Therefore, H I gas, i.e., the [C II] emission, should be distributed in a way that it is covering the ionized gas partially if an efficient Lyman continuum escape is the case. On the other hand, the UV continuum in the J band is unresolved due to the limited spatial resolution of the ground-based observation (Inoue et al. 2016). The weak dust emission is in line with the large Lyα EW and efficient escape of the Lyman continuum from SXDF-NB1006-2.

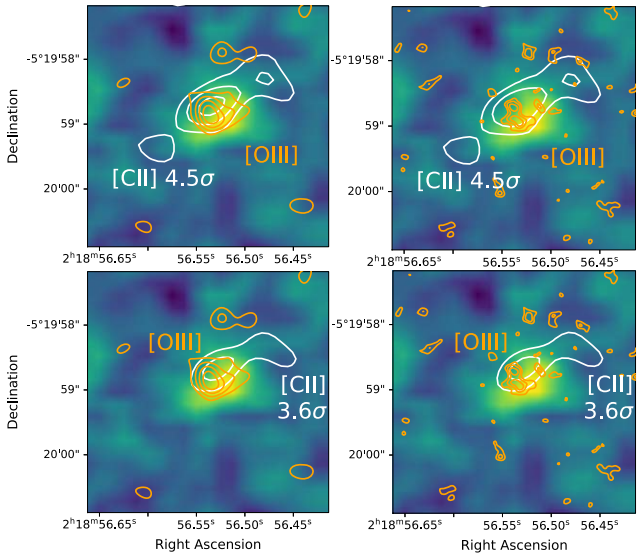

**Figure 10.** Spacial distribution of the Lyα, [O III] 88 μm, and [C II] 158 μm emission. The background of all panels is the Lyα emission measured by the Subaru telescope. The orange and white lines indicate $+2\sigma$, $+3\sigma$, $+4\sigma$, and $+5\sigma$ contours of [O III] and [C II] emission, respectively. The top-left panel shows the contours of the 4.5σ [C II] emission and 5.9σ [O III] emission. The top-right panel shows the contours of the 4.5σ [C II] emission and clumpy ∼3–4σ [O III] emission. The bottom-left panel shows the contours of the 3.6σ [C II] emission and 5.9σ [O III] emission. The bottom-right panel shows the contours of the 3.6σ [C II] emission and clumpy ∼3–4σ [O III] emission.

## 5. Conclusion

In this work, we analysed two [O III] ALMA observations and three [C II] ALMA observations to study the [O III] 88 μm, [C II] 158 μm, and dust continuum emission from a $z = 7.212$ galaxy, SXDF-NB1006-2. We discussed the [O III]/[C II] luminosity ratios and the $L_{[C II]}$−SFR relation updated by our results. We summarize our conclusions as follows:

1. With the high angular resolution and sensitivity observations, the [O III] 88 μm emission shows a clumpy structure, having two or three components with marginal >3σ detections. We also performed detailed velocity integrals to optimize the detected S/Ns. The scales of the clumps are 0.32–0.85 kpc. However, it is possible that a smooth disk can look clumpy in images obtained from high angular resolution interferometric observations (Gullberg et al. 2018), and we cannot rule out this possibility for our clumpy [O III] detection. For the entire [O III] 88 μm emission of SXDF-NB1006-2, we finally obtained a 5.9σ detection after combining all the [O III] data and performing a 0″.3 uv-taper. The line luminosity increases by a factor of 1.25 compared with Inoue et al. (2016), but is still consistent within the uncertainty.

2. We found that the [C II] 158 μm signal is unstable among the three ALMA data sets: there are ∼3–4σ marginal signals in two data sets, but nothing in the other data set. After combining all the [C II] data sets, we obtained a marginal detection of 4.5σ. However, the robustness of this detection is low, because after we integrated without one velocity bin at the integral edge, the S/N decreases to 3.6σ, and the luminosity decreases by a factor of ∼2. The [C II] 158 μm emission of SXDF-NB1006-2 may also has subcomponents on a larger scale different from the [O III] case.

3. The dust continuum emission remains undetected, and we obtained more stringent constraints on the total IR luminosity than the previous study. This indicates little dust content in SXDF-NB1006-2, consistent with the bright Lyα emission.

4. The observed [O III]/[C II] luminosity ratio is $10.2 \pm 4.7$ and $20 \pm 12$ for the 4.5σ and 3.6σ [C II] detections, respectively. We also take into account the SBD effect. Then, the [O III]/[C II] luminosity ratio reduced to $6.1 \pm 3.5$ and $9.6 \pm 6.1$ for the 4.5σ and 3.6σ [C II] detections, respectively. All of these ratios are consistent with those of local dwarf galaxies within the large dispersion but close to the highest boundary.

5. We compared our results with the $L_{[C II]}$−SFR relations derived from local H II/starburst galaxies, $4 < z < 6$ starforming galaxies of the ALPINE survey, and $z \sim 6$–9 galaxies with modified [C II] detections and corrected for the SBD effect (De Looze et al. 2014; Carniani et al. 2020; Schaerer et al. 2020). The data points of SXDF-NB1006-2 are located below these relations, indicating a strong [C II] deficit given its SED-based SFR. However, there are large uncertainties in the measurements of the SFR from SED fitting, and we hope future JWST observations can help solve this problem.

We thank the anonymous referee's precious comments and suggestions. Y.F., A.K.I., and Y.S. are supported by NAOJ ALMA Scientific Research grant No. 2020-16B. K.K. and Y.T. were supported by JSPS KAKENHI grant No. 17H06130. T.H. was supported by Leading Initiative for Excellent Young Researchers, MEXT, Japan (HJH02007) and by JSPS KAKENHI grant Nos. 20K22358 and 22H01258. T.O. acknowledges JSPS KAKENHI grant Nos. 21H04496, 20H05861, and 19H01931.

This paper makes use of the following ALMA data: ADS/JAO.ALMA#2012.1.00374.S, 2013.1.01010.S, 2013.A.00021.S, 2015.A.00018.S, and 2019.1.01634.L. ALMA is a partnership of ESO (representing its member states), NSF (USA), and NINS (Japan), together with NRC (Canada), MOST and ASIAA (Taiwan), and KASI (Republic of Korea), in cooperation with the Republic of Chile. The Joint ALMA Observatory is operated by ESO, AUI/NRAO, and NAOJ.

## ORCID iDs

Yi W. Ren https://orcid.org/0000-0002-6510-5028
Yoshinobu Fudamoto https://orcid.org/0000-0001-7440-8832
Akio K. Inoue https://orcid.org/0000-0002-7779-8677
Yuma Sugahara https://orcid.org/0000-0001-6958-7856
Yoichi Tamura https://orcid.org/0000-0003-4807-8117
Hiroshi Matsuo https://orcid.org/0000-0003-3278-2484
Kotaro Kohno https://orcid.org/0000-0002-4052-2394
Hideki Umehata https://orcid.org/0000-0003-1937-0573
Takuya Hashimoto https://orcid.org/0000-0002-0898-4038
Rychard J. Bouwens https://orcid.org/0000-0002-4989-2471
Renske Smit https://orcid.org/0000-0001-8034-7802
Nobunari Kashikawa https://orcid.org/0000-0003-3954-4219
Takashi Okamoto https://orcid.org/0000-0003-0137-2490
Ikkoh Shimizu https://orcid.org/0000-0001-5686-8368